# Chemical ordering in $Pd_{81}Ge_{19}$ metallic glass studied by reverse Monte-Carlo modelling of XRD, ND and EXAFS experimental data

Ildikó Pethes[1,*], Ivan Kaban[2], Mihai Stoica[2], Brigitte Beuneu[3] and Pál Jóvári[1]

[1]Wigner Research Centre for Physics, Hungarian Academy of Sciences, H-1525 Budapest, POB 49, Hungary

[2]IFW Dresden, Institute for Complex Materials, Helmholtzstr. 20, 01069 Dresden, Germany

[3]Laboratoire Léon Brillouin, CEA-Saclay 91191 Gif sur Yvette Cedex, France

E-mail: pethes.ildiko@wigner.mta.hu

**Abstract.** $Pd_{81}Ge_{19}$ metallic glass was investigated by neutron diffraction, X-ray diffraction and extended X-ray absorption fine structure spectroscopy (EXAFS) at the Ge K-edge. Large scale structural models were obtained by fitting the three measurements simultaneously in the framework of the reverse Monte Carlo simulation technique. It was found that the experimental data sets can be adequately fitted without Ge-Ge nearest neighbours. Mean Pd-Pd and Pd-Ge distances are 2.80±0.02 Å and 2.50±0.02 Å, respectively. The total average coordination number of Pd is 12.1±0.5 while Ge is surrounded by 10.6±1.1 Pd atoms. The coordination numbers calculated from partial pair correlation functions were compared to those obtained by Voronoi tessellation method. It was found that the latter technique overestimates the number of nearest neighbours by about 20% due to the significant contribution of distant pairs.

---

[*]Corresponding author.

## 1. Introduction

Though metallic glasses were discovered more than 50 years ago, their structure is still in the focus of the interest of many recent publications [1, 2]. The experimental and computational facilities evolved greatly in this period providing opportunity to construct realistic three-dimensional model of amorphous structures.

In a glassy state of matter the constituent atoms have no exact positions (in the sense of atomic positions in a unit cell). Therefore the goal of structural studies is to extract statistical information about the positions of atoms in a glass. The term 'structure' used in the context of amorphous materials may either mean topological structure (arrangement of the atoms according to the constraints of efficient space filling), chemical order (preferred neighbours around the different components forming the alloy) or short or medium range structure.

The tools of the structural characterization are experimental techniques such as X-ray or neutron diffraction (XRD, ND), X-ray absorption fine structure (EXAFS), transmission electron microscopy, nuclear magnetic resonance, and computer simulation methods such as molecular dynamics (MD) simulations, ab initio MD and Reverse Monte Carlo (RMC) modelling [3].

In this work we investigate the atomic structure of Pd$_{81}$Ge$_{19}$ metallic glass. It belongs to the class of transition-metal–metalloid (TM–M) alloys for which the glass formation was established for the first time in 1960 [4, 5]. Even early models suggested chemical ordering in these glasses: the atoms of the minority metalloid components avoid the vicinity of each other [6, 7]. This theory has been verified by the study of Hayes and co-workers [8], who investigated Pd$_{78}$Ge$_{22}$ and Pd$_{80}$Ge$_{20}$ amorphous alloys by EXAFS measurements at the Ge K-edge. They found that around each Ge atom there are 8.6±0.5 Pd and considerably less than one Ge nearest neighbour. (Similar coordination numbers were found earlier in Pd$_{84}$Si$_{16}$ by X-ray and neutron diffraction measurements [9].) Trigonal prismatic local structure with 9 Pd around each Ge atoms in Pd$_{80}$Ge$_{20}$ was proposed by EXAFS and X-ray absorption near-edge structure (XANES) investigations [10]. As we are not aware of any recent experimental studies on Pd-Ge glasses it may be interesting to have a new look at this old problem using more advanced techniques.

## 2. Experimental

About 10 g of Pd$_{81}$Ge$_{19}$ master alloy was prepared by arc-melting in Ar atmosphere from pure Pd (99.95%) and Ge (99.999%). The alloy was re-melted several times to insure homogeneity. Glassy ribbons were obtained by melt spinning under Ar on a copper wheel rotating with a tangential velocity of 38 m·s$^{-1}$. Pd$_{81}$Ge$_{19}$ ingot was melted and heated in a quartz tube up to about 1100 °C before ejection.

High-energy XRD measurement was performed at the BW5 beamline [11] at HASYLAB (DESY, Hamburg) in transmission geometry. The energy of the incident beam was 84.98 keV. The size of the incident beam was $1 \times 2$ mm$^2$. The scattered intensity was recorded by a Ge solid-state detector. The raw experimental data were corrected for detector deadtime, background, polarization, absorption and variation in detector solid angle [12].The neutron diffraction measurements were carried out with the 7C2 diffractometer at the Léon Brillouin Laboratory (CEA-CNRS Saclay, France). The wavelength of the incident radiation was 0.73 Å. Pieces of amorphous ribbons were placed into a thin-walled (0.1 mm) vanadium container. The raw data were corrected for the detector efficiency, background scattering, absorption, multiple and incoherent scattering.

The EXAFS measurements at the Pd K-absorption edge were carried out at the beamline X1 at DESY, Hamburg. The spectra were collected in transmission mode using fixed exit double-crystal Si(311). The intensities before and after the sample as well as after the reference sample were recorded by three ionization chambers (IOC) filled with proper quantities of Ar and Kr with pressures depending on the edge energy. The EXAFS-modulations were extracted from the recorded spectra using the VIPER program [13].

## 3. Simulation and calculation methods

### 3.1. Reverse Monte Carlo simulations

The X-ray or neutron diffraction total structure factors are related to the partial structure factors $S_{ij}(Q)$ by Eq. (1):

$$S(Q) = \sum_{i \leq j} w_{ij}^{X,N}(Q) S_{ij}(Q). \tag{1}$$

The scattering weights for X-ray diffraction are:

$$w_{ij}^X(Q) = (2 - \delta_{ij}) \frac{c_i c_j f_i(Q) f_j(Q)}{\left(\sum_i c_i f_i(Q)\right)^2} \tag{2}$$

and for neutron diffraction they are:

$$w_{ij}^N = (2 - \delta_{ij}) \frac{c_i c_j b_i b_j}{\left(\sum_i c_i b_i\right)^2}. \tag{3}$$

$Q$ is the amplitude of the scattering vector, $\delta_{ij}$ is the Kronecker delta, $c_i$ is the atomic concentration, $f_i(Q)$ is the atomic form factor, $b_i$ is he coherent neutron scattering length.

For Pd$_{81}$Ge$_{19}$ the neutron diffraction weight factors are $w_{PdPd} = 0.5697$, $w_{PdGe} = 0.3702$ and $w_{GeGe} = 0.0601$. The XRD weight factors are shown in figure 1.

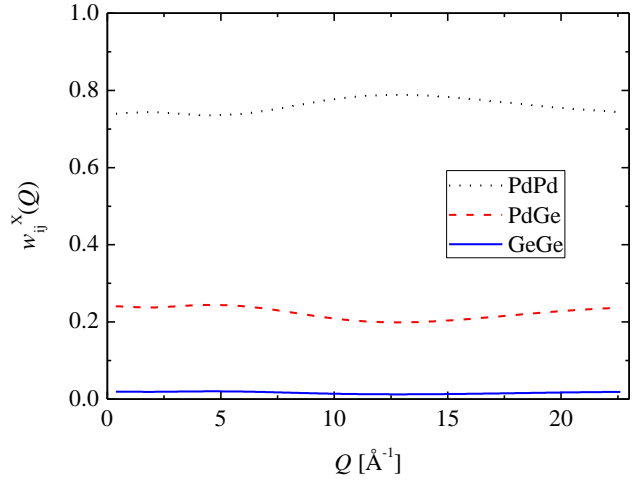

**Figure 1.** XRD weight factors for Pd$_{81}$Ge$_{19}$ alloy.

The relationship between the Faber-Ziman $S_{ij}(Q)$ and the partial pair correlation functions is given by:

$$g_{ij}(r) - 1 = \frac{1}{2\pi^2 \rho_0 r} \int_0^\infty Q[S_{ij}(Q) - 1] \sin(Qr) dQ \tag{4}$$

or

$$S_{ij}(Q) - 1 = \frac{4\pi \rho_0}{Q} \int_0^\infty r[g_{ij}(r) - 1] \sin(Qr) dr, \tag{5}$$

where $\rho_0$ is the average number density.

The experimental x-ray absorption coefficient is converted into the EXAFS signal $\chi(k)$ (as a function of the wavenumber $k$ of the photoelectron $k = \sqrt{2m_e(E - E_0)}/\hbar$, $m_e$ is the electron mass, $\hbar$ is the reduced Planck constant), which is related to the partial pair correlation functions by:

$$\chi_i(k) = \sum_j 4\pi \rho_0 c_j \int_0^R r^2 \gamma_{ij}(k,r) g_{ij}(r) dr. \tag{6}$$

Here $i$ refers to the absorber atom, $\gamma_{ij}(k,r)$ is the photoelectron backscattering matrix, which gives the $k$-space contribution of a $j$-type backscatterer at distance $r$ from the absorber atom. Elements of $\gamma_{ij}(k,r)$ matrix were calculated for each $i$-$j$ pair by the FEFF8.4 program [14].

The measured ND and XRD structure factors and the $k^3$ weighted $\chi(k)$ EXAFS curve are shown in figure 2.

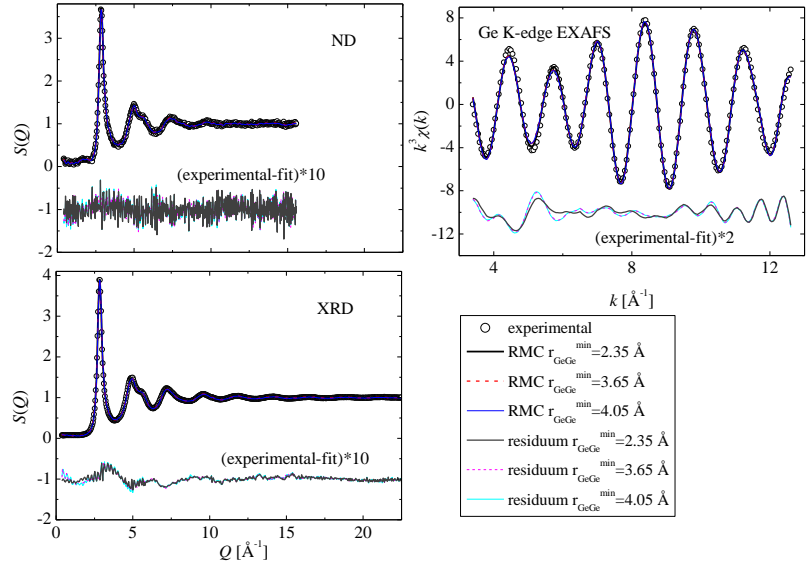

**Figure 2.** Experimental ND and XRD structure factors and $k^3$ weighted $\chi(k)$ EXAFS curve of $Pd_{81}Ge_{19}$ glass (symbols), and model curves (lines) obtained by RMC simulation. Results of the three different simulation runs in which the Ge-Ge minimum interatomic distances were 2.35 Å (thick solid line), 3.65 Å (dashed line, final model) and 4.05 Å (thin solid line). The residuals are shifted by -1 for the diffraction and by -10 for EXAFS.

In RMC modelling partial pair correlation functions, $S(Q)$ and $\chi(k)$ functions are calculated from the generated atomic coordinates. During the simulation particles are moved around randomly to minimize the differences between model ($S_{mod}(Q)$, $\chi_{mod}(k)$) and experimental ($S_{exp}(Q)$, $\chi_{exp}(k)$) curves. The particle configurations obtained by this procedure are compatible with all of the experimental data sets (within their experimental errors).

The RMC++ [15] code was used in this study. Simulation boxes contained 10000 atoms in test runs and 30000 particles in the final runs, which were analysed further. The atomic number density of 0.06202 Å$^{-3}$ was calculated from the reported mass density of 10.3±0.3 g·cm$^{-3}$ for $Pd_{80}Ge_{20}$ metallic glass [16]. The minimum interatomic distances were 2.3 Å and 2.2 Å for Pd-Pd and Pd-Ge pairs. Dedicated simulation runs were carried out to determine the necessity of Ge-Ge nearest neighbours in which the minimum interatomic distance for Ge-Ge pairs was varied between 2.35 Å and 4.45 Å. During the final runs the Ge-Ge minimum interatomic distance was 3.65 Å. Initial configurations were obtained by placing atoms into the box randomly and moving them around to satisfy the minimum interatomic distance requirements. The $\sigma$ parameters used to calculate the RMC cost function [3] were reduced in three steps to the final values of $2\times10^{-3}$ for the diffraction data sets and $2\times10^{-5}$ for the EXAFS data. The number of accepted moves was around $1\text{-}2 \times 10^7$.

*3.2. Voronoi tessellation calculations*

Voronoi tessellation [17] calculations were performed by the program Voro++ [18]. Although the Pd and Ge atoms have similar atomic sizes, the radical Voronoi tessellation method was used to avoid the discrepancies different atomic sizes can cause [19]. The final model of 30000 atoms obtained by the RMC method was investigated. To reduce the statistical errors results of 5 configurations were averaged during Voronoi cell analysis.

**4. Results and discussion**

*4.1. Investigation of Ge-Ge nearest neighbours*

According to chemical ordering the number of Ge-Ge nearest neighbours is expected to be negligible. The validity of this assumption was investigated by two series of dedicated runs eliminating Ge-Ge bonding in two different ways.

In the first series, the Ge-Ge minimum interatomic distance ($r_{GeGe}^{min}$) was increased in small increments between 2.35 Å and 4.45 Å. The possible mean nearest neighbour distance is expected in the region of the covalent and metallic diameters, which is around 2.4-2.8 Å. The compatibility of the experimental data sets with these minimum distance values was examined through the quality of the fits, which was monitored through their *R*-factors. The *R*-factor of a given data set (diffraction or EXAFS) is defined by:

$$R = \frac{\sqrt{\sum_i [S_{mod}(Q_i) - S_{exp}(Q_i)]^2}}{\sqrt{\sum_i S_{exp}^2(Q_i)}}, \qquad (7)$$

where $S_{mod}$ and $S_{exp}$ are the model and experimental structure factors or $\chi(k)$ curves, respectively. The *R*-factors of the different models were compared to that of the model with the lowest Ge-Ge minimum interatomic distance ($r_{GeGe}^{min}$ = 2.35 Å) (the reference model).

The values of a relative *R*-factor (*R*(model)/*R*(reference model)) are shown as a function of the minimum Ge-Ge distance in figure 3. The model curves obtained for $r_{GeGe}^{min}$ = 2.35 Å and $r_{GeGe}^{min}$ = 4.05 Å are shown in figure 2. The values of the *R*-factors scarcely changed up to $r_{GeGe}^{min} \approx 4$ Å. Although the weight of the Ge-Ge partial is small in the two diffraction experiments, the higher $r_{GeGe}^{min}$ values can cause changes in the $g_{PdGe}(r)$ curve also, which has higher weight in these measurements. As the Ge-Ge and Ge-Pd backscattering factors oscillate in opposite phase (see figure 4) the EXAFS curve is the most sensitive to the presence of Ge-Ge pairs, but the quality of the EXAFS fit remained as good for $r_{GeGe}^{min}$ = 4.05 Å as for lower minimum interatomic distances.

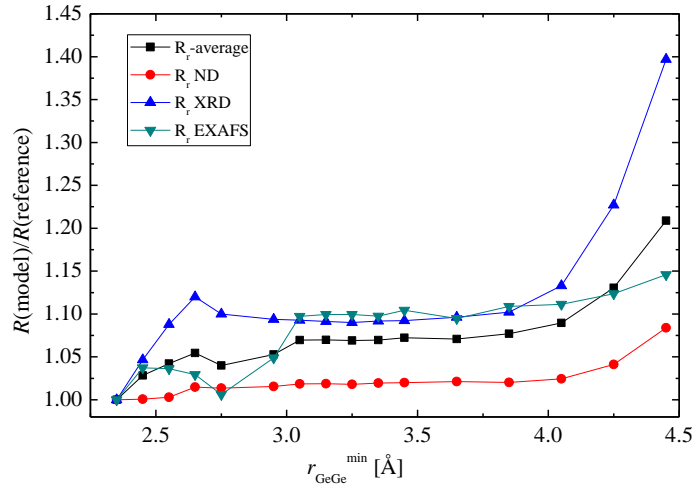

**Figure 3.** Relative *R*-factors (*R*(model)/*R*(reference model)) of the data sets by using different Ge-Ge minimum interatomic distances during the simulation runs. $r_{GeGe}^{min}$ = 2.35 Å for the reference model. The average of the three relative *R*-factors is also shown.

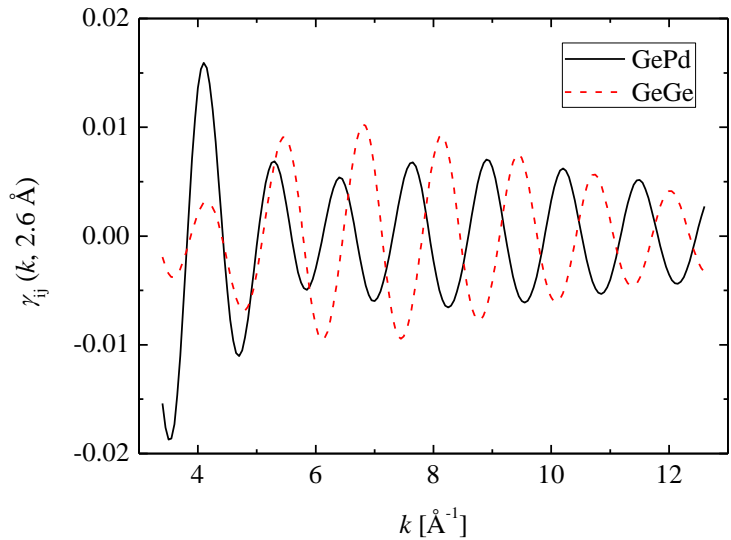

**Figure 4.** Values of the backscattering matrix $\gamma_{ij}(k,r)$ at $r$ = 2.6 Å. (solid line: *ij* = GePd, dashes: *ij*=GeGe)

In the second series of test runs the Ge-Ge minimum interatomic distance was kept in 2.35 Å, but the number of Ge-Ge pairs was reduced by constraining the number of the Ge-Ge pairs between 2.35 Å ≤ $r_{GeGe}$ ≤ 3.60 Å. The upper limit was chosen on the basis of the unconstrained (reference) model: this value is between the possible nearest neighbour distance (first peak in *g*(*r*)) and the second coordination shell.

The average number of Ge-Ge pairs (average coordination number) for the above defined *r* region in the unconstrained simulation run was $N_{GeGe}$ = 1.8. This value was reduced to 0 in 6 steps. The obtained $g_{GeGe}(r)$ curves are shown in figure 5. The relative $R$-factors of the data sets are shown in table 1.

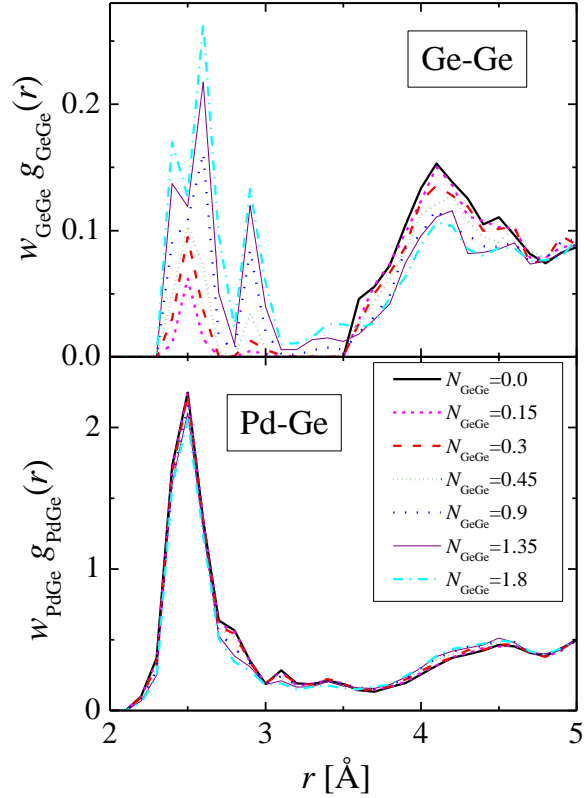

**Figure 5.** $w^N$ weighted $g_{GeGe}(r)$ and $g_{PdGe}(r)$ partial pair correlation functions are obtained by RMC modelling using average coordination constraints for the $N_{GeGe}$ coordination number.

**Table 1.** $R$-factors of the data sets obtained by RMC modelling using average coordination constraints for $N_{GeGe}$. The average of the three relative $R$-factors, $R_c$ is also shown (the relative $R$-factor is $R$(model)/$R$(reference model)). The reference model (denoted with 'free') was obtained by unconstrained simulation.

| $N_{GeGe}$ | $R$-ND | $R$-XRD | $R$-EXAFS | $R_c$ |
|---|---|---|---|---|
| 0 | 0.0469 | 0.0227 | 0.0865 | 1.08 |
| 0.15 | 0.0465 | 0.0220 | 0.0843 | 1.05 |
| 0.30 | 0.0464 | 0.0214 | 0.0825 | 1.03 |
| 0.45 | 0.0462 | 0.0212 | 0.0820 | 1.03 |
| 0.90 | 0.0461 | 0.0209 | 0.0800 | 1.01 |
| 1.35 | 0.0460 | 0.0206 | 0.0788 | 1.00 |
| 1.80 ('free') | 0.0459 | 0.0205 | 0.0785 | 1.00 |

The quality of the fits only slightly decreases as the GeGe coordination number is getting smaller. The average of the three relative *R*-factors is below 1.1 at the lowest $N_{GeGe} = 0$ value. The $g_{GeGe}(r)$ curves for $N_{GeGe} > 0.3$ have 3-4 peaks in the 2.35 Å $\leq r \leq$ 3.6 Å region, which seems to be unrealistic, suggesting that $N_{GeGe} \leq 0.3$.

Both series of test runs (changing $r_{GeGe}^{min}$ without coordination constraint and changing $N_{GeGe}$ without changing $r_{GeGe}^{min}$) show that the number of Ge-Ge nearest neighbours is negligible (or at least below the sensitivity of our method). Thus for further investigations (for the final simulation run) $r_{GeGe}^{min}$ was chosen to be 3.65 Å. The partial pair correlation functions of the model with the final minimum interatomic distances are shown in figure 6, the fits of data sets are presented in figure 2.

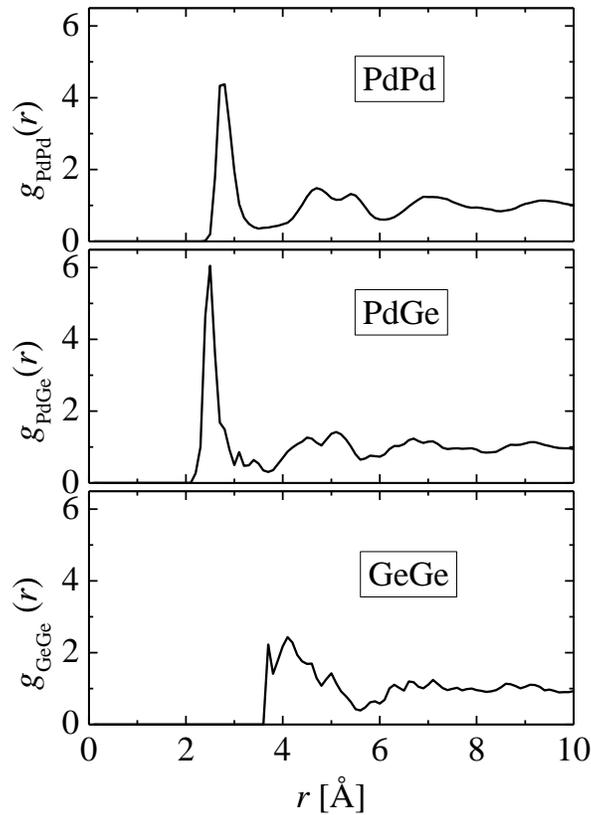

**Figure 6.** Partial pair correlation functions of the final model obtained by RMC simulation. During this simulation run coordination constraints were not used, and the Ge-Ge minimum interatomic distance was 3.65 Å.

*4.2. Coordination numbers*

The $N_{ij}$ coordination number is the average number of *j* type atoms in the nearest neighbour shell of a given *i*-type central atom. For crystalline materials, the term 'nearest neighbour' is well-defined, while for amorphous substances it is rather depending on the method of calculation. According to the 'distance cutoff method' all atoms contributing to the first peak of the partial pair correlation function up to the 'cutoff distance' ($r_{max}$) are considered as nearest neighbour. The 'cutoff distance' is generally the first minimum of the pair correlation function. The coordination number then can be given as

$$N_{ij} = 4\pi\rho_0 c_j \int_0^{r_{max}} g_{ij} r^2 dr . \tag{8}$$

The $r_{max}$ value is well-defined in several covalent glasses, where the first and second coordination spheres are noticeably separated and $g_{ij}(r_{max}) \approx 0$. However in metallic glasses the value of $r_{max}$ is not so obvious as $g_{ij}(r)$ usually does not reach zero.

The Pd-Pd partial pair correlation function $g_{PdPd}(r)$ has an unambiguous peak around at a mean nearest neighbour distance $r_{PdPd}$ = 2.80 (± 0.02) Å, followed by a minimum at $r$ = 3.5 Å. It seems to be adequate to choose $r_{max}$ = 3.5 Å. However, as $g_{PdPd}$(3.5 Å) is not equal to 0, a slightly smaller or higher $r_{max}$ value would result in different $N_{PdPd}$ values.

The situation is already more difficult for the Pd-Ge partial. It has a peak around $r_{PdGe}$ = 2.50 (± 0.02) Å together with a shoulder (at 2.7-2.8 Å), and followed by smaller peaks at 3.1 Å and 3.4 Å. The minimum of $g_{PdGe}(r)$ is around $r_{max}$=3.6 Å.

The coordination numbers obtained by these $r_{max}$ values are shown in table 2. The Ge-Ge nearest neighbours can be eliminated, as was shown in the previous section. Thus, for Ge-Ge pairs $r_{max} = r_{GeGe}^{min}$ (minimum interatomic distance) = 3.65 Å and $N_{GeGe}$ = 0. The total coordination numbers with the above mentioned cutoff distances are $N_{Pd}$ = 12.1±0.5 and $N_{Ge}$ = 10.6±1.1.

The $g_{ij}(r)$ curves and the values of the coordination numbers were monitored in the series of simulation runs presented in section 4.1, where $r_{GeGe}^{min}$ and $N_{GeGe}$ were changed systematically. The amplitudes of the main peak and the small peaks as well as the shoulder in the $g_{PdGe}(r)$ curves are slightly sensitive to the value of $r_{GeGe}^{min}$ and $N_{GeGe}$ (see figure 5). In contrast to this, the position of the minimum in the $g_{PdGe}(r)$ remains around $r$=3.6 Å. The decrease of $N_{GeGe}$ is largely compensated by the increase of $N_{GePd}$ keeping thus the total average coordination number of Ge always around 10.6-10.7. The height of the first peak of the $g_{PdPd}$ slightly decreases as $N_{GeGe}$ is reduced; the position of the minimum however remains at 3.5 Å, which brings about a decrease of $N_{PdPd}$. It was found that the $N_{Pd}$ total coordination number is always around 12-12.1, regardless of the value of $N_{GeGe}$.

Knowing the total coordination number one can calculate the random partial coordination numbers: the coordination numbers of a configuration in which the Ge and Pd atoms are placed randomly. They can be obtained as follows:

$$N_{ij}^{r} = \frac{N_i c_j N_j}{\sum_k c_k N_k}, \tag{9}$$

where $N_i$ and $c_i$ are the total coordination number and concentration of the $i^{th}$ component. Taking into account that $N_{Pd}$ = 12.1 and $N_{Ge}$ = 10.6, the random partial coordination numbers are: $N^r_{PdPd}$ = 10.04, $N^r_{PdGe}$ =2.06, $N^r_{GePd}$ = 8.79, and $N^r_{GeGe}$ = 1.81. These values are very close to the values obtained for the unconstrained run with $r_{GeGe}^{min}$ = 2.35 Å.

Table 2. Coordination numbers obtained by different calculation methods.

|  | 'cutoff-distance method' | Voronoi tessellation method 'cutoff area-ratio' 0% | Voronoi tessellation method 'cutoff area-ratio' 1% |
|---|---|---|---|
| $N_{PdPd}$ | 9.63 | 11.60 | 10.43 |
| $N_{PdGe}$ | 2.48 | 2.87 | 2.58 |
| $N_{GePd}$ | 10.58 | 12.31 | 11.00 |
| $N_{GeGe}$ | 0.0 | 0.88 | 0.40 |
| $N_{Pd}$ | 12.11 | 14.49 | 13.01 |
| $N_{Ge}$ | 10.58 | 13.18 | 11.40 |

Another widely used method for the determination of the coordination number is the Voronoi tessellation method [17]. In this method the 3-dimensional space is divided into cells centered by atoms: a plane is drawn to bisect each line connecting the central atom and one of the neighbouring atoms, and the cell defined by all such planes is called a Voronoi cell. The Voronoi cell (polyhedron) of a central atom encloses the part of the space that is closer to the central atom than any other atoms. This method can be used to determine the coordination number without a cutoff distance: those atoms sharing a common cell surface are considered to be nearest neighbours.

The total coordination numbers of Pd and Ge, obtained by using the Voronoi tessellation method for the final configuration are 14.5 and 13.2 (see table 2), which are considerably higher than the values obtained by the 'distance cutoff method' from the partial pair correlation functions. The distribution of the nearest neighbour distances of the 'Voronoi nearest neighbours' (the neighbours obtained by the Voronoi tessellation method) is shown in figure 7a. There are several pairs which have much higher distances than the previously applied $r_{max}$ values. Some of these 'nearest neighbours' are as far apart as 5-6 Å, which is twice the mean nearest neighbour distance of the total pairs. It raises the question whether these pairs can be regarded as nearest neighbours.

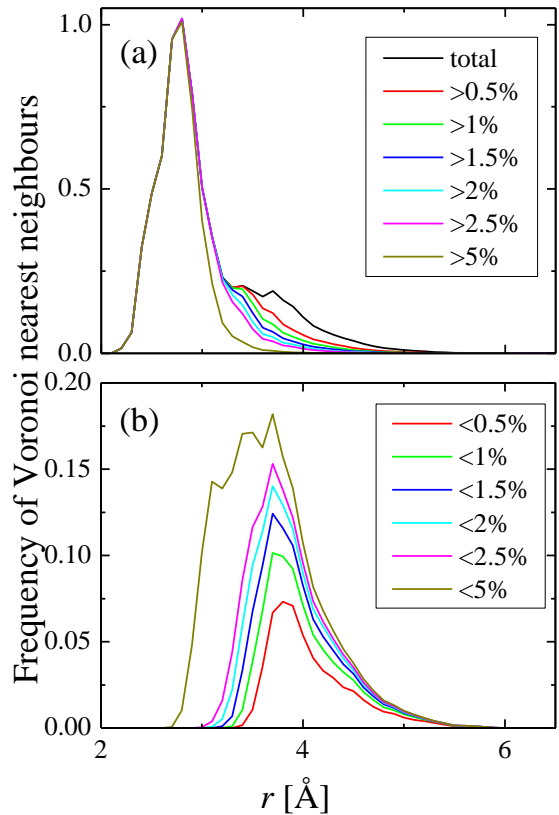

**Figure 7.** Distance-distributions of Voronoi cells sharing a common face with area smaller (a) or bigger (b) than the 'cutoff area-ratio' percent of the total surface of the Voronoi cells.

Obviously, the contribution to the surface of the Voronoi cell of the pairs with different distances is not uniform: distant atoms usually have small common cell surface. Neighbours sharing faces area of which is less than 1 % of the total cell surface, have a mean distance as high as 3.8 Å (see figure 7b). However, almost all the pairs that share a common face with an area higher than 5 % of the total cell surface are closer to each other than 3.75 Å (figure 7a).

This inspires the usage of a properly chosen 'cutoff area-ratio' to eliminate the majority of distant 'nearest neighbours'. Unfortunately, this again is not a well-defined parameter.

Removing the small faces was used earlier by Sheng and co-workers [2]; they excluded the faces area of which were less than 1% of the total surface of the polyhedron. They pointed out that the 'degeneracy problem and the effects of thermal vibration are minimized' with this procedure.

The next step is to combine the two methods and find adequate $r_{max}$ and 'cutoff area-ratio' parameters using them together. At first Voronoi coordination numbers are calculated taking into account only pairs with common face area higher than the 'cutoff area-ratio'. Then $r_{max}$ is calculated for this coordination number. $r_{max}$ values appertain to different 'cutoff area-ratio' parameters are presented in table 3 and some of them are shown together with the partial pair correlation functions in figure 8.

**Table 3.** Different 'cutoff area-ratio' parameters and corresponding $r_{max}$ values.

| 'cutoff area-ratio' [%] | $r_{max}$(PdPd) [Å] | $r_{max}$(PdGe) [Å] | $r_{max}$(GeGe) [Å] |
|---|---|---|---|
| 0 | 4.03 | 4.03 | 3.88 |
| 0.5 | 3.845 | 3.89 | 3.765 |
| 1 | 3.75 | 3.78 | 3.735 |
| 1.5 | 3.66 | 3.65 | 3.71 |
| 2 | 3.58 | 3.55 | 3.695 |

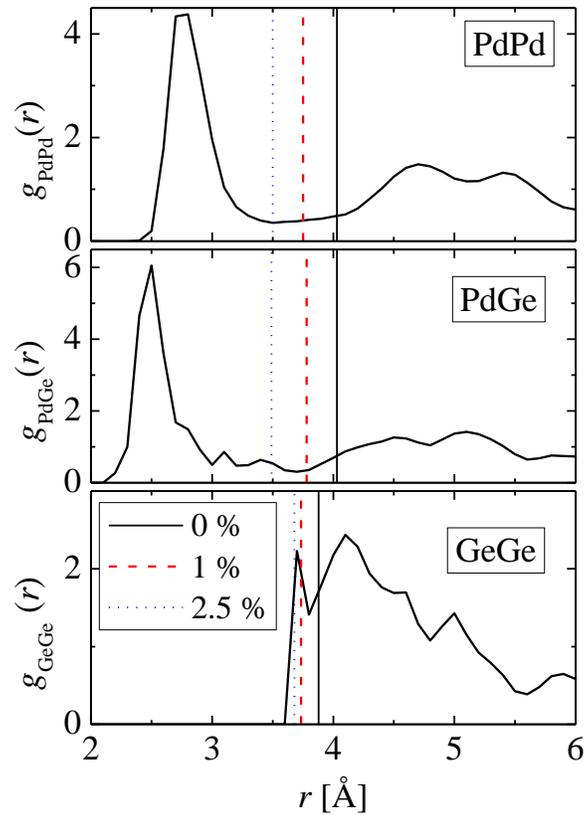

**Figure 8.** $r_{max}$ values obtained by various 'cutoff area ratio' parameters (solid line: 0%, dashes: 1%, dots: 2.5%)

To obtain the same coordination number in the 'cutoff distance method' as by the Voronoi tessellation method without any excluded pair, $r_{max}$ should be about 3.9 Å (for GeGe) and 4.0 Å (for PdGe and PdPd). These values seem to be too high; the $g_{GeGe}$ and $g_{PdGe}$ curves definitely start to increase at lower $r$ values.

The $r_{max}$ values selected at the 'distance cutoff method' require a 'cutoff area-ratio' parameter about 2.5%. However, by this choice a lot of 'real' nearest neighbours (with distance lower than 3.5 Å) are removed, too.

The compromise between the two methods can be halfway: exclude the pairs with common face's area less than 1 % of the total cell surface, which is associated with $r_{max}$ about 3.75 Å. Coordination numbers for this selection are shown in table 2.

For the investigated configuration about 11% of the 'nearest neighbour' pairs have common face area less than 1% of the total cell-surface. Excluding these pairs causes great difference not only in the total coordination numbers but in the distribution of the Voronoi indexes as well (see next section).

The distributions of the total and average coordination numbers were also investigated. The calculations were performed for three sets of the cutoff distance ($r_{max}$) values: (1) those of the 'distance cutoff method' ($r_{max}$ = 3.5 Å, 3.6 Å and 3.6 Å for Pd-Pd, Pd-Ge and Ge-Ge, respectively), (2) the $r_{max}$ values belonging to the 1 % 'cutoff area-ratio' (3.75 Å, 3.78 Å and 3.735 Å), and (3) the $r_{max}$ values belonging to the 0% 'cutoff area-ratio' (4.03 Å, 4.03 Å, 3.88 Å). The distributions are shown in figure 9. The shapes of the distribution curves are similar for the different cases. The peaks are shifting towards higher values as the $r_{max}$ values are raised. The distributions of the total coordination numbers are slightly wider and the peaks are somewhat lower for the higher $r_{max}$ values.

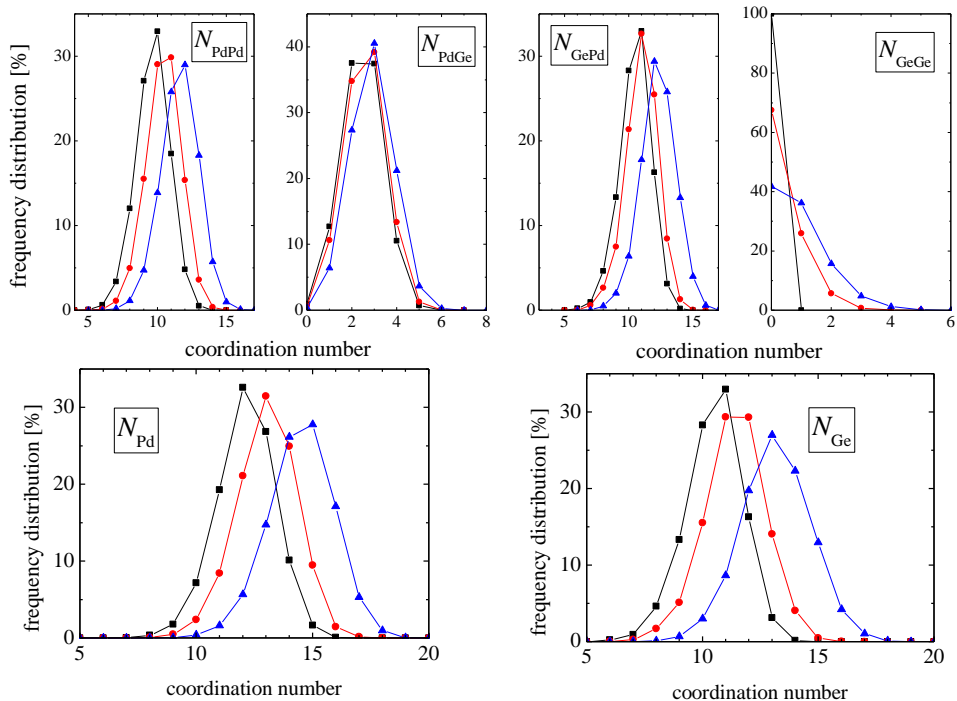

**Figure 9.** Frequency distributions of the coordination numbers obtained for different $r_{max}$ values. Squares: $r_{max}$(PdPd)=3.5Å, $r_{max}$(PdGe)=3.6Å and $r_{max}$(GeGe)=3.6Å. Circles: $r_{max}$(PdPd)=3.75Å, $r_{max}$(PdGe)=3.78Å and $r_{max}$(GeGe)=3.735 Å. Triangles: $r_{max}$(PdPd)=4.03Å, $r_{max}$(PdGe)=4.03Å, $r_{max}$(GeGe)=3.88Å. The lines

are just guides to the eye.

*4.3. Voronoi cell analysis*

Voronoi cells can be labelled by the Voronoi index, $< n_3, n_4, n_5, n_6 >$, where $n_i$ represents the number of $i$-edged faces of the polyhedron ($\sum n_i$ =coordination number). There are usually several hundred types of different polyhedra in the metallic glasses. The frequency distribution of the Voronoi polyhedra was investigated for the final model containing 30000 atoms. As the first coordination shell of germanium atoms differs from that of the palladium (the coordination number of the latter is significantly higher) the Ge-centered and Pd-centered polyhedra were investigated separately.

The most prominent Voronoi polyhedra types in the studied $Pd_{81}Ge_{19}$ glass are presented in tables 4 and 5.

**Table 4.** Frequency distribution of the Voronoi polyhedron around Ge atoms. Case (a) denotes the distribution of the original Voronoi cell indexes (without excluding small faces), values of case (b) were calculated with exclusion of faces with are less than 1%.

| Voronoi-index | Frequency (%) | |
|---|---|---|
| $< n_3, n_4, n_5, n_6 >$ | case (a) | case (b) |
| <0,2,8,2> | 6.04 | 3.07 |
| <0,2,8,1> | 5.36 | 3.31 |
| <0,3,6,4> | 4.61 | 1.07 |
| <0,3,6,3> | 3.67 | 1.45 |
| <1,2,6,2> | 2.12 | 0.12 |
| <0,2,6,3> | 0 | 2.63 |
| <0,1,8,2> | 0 | 2.61 |
| <0,2,6,4> | 0 | 2.38 |
| <0,2,6,2> | 0 | 2.31 |
| <0,1,8,1> | 0 | 2.25 |

**Table 5.** Frequency distribution of the Voronoi polyhedron around Pd atoms. Case (a) denotes the distribution of the original Voronoi cell indexes (without excluding small faces), values of case (b) were calculated with exclusion of faces with are less than 1%.

| Voronoi-index $< n_3, n_4, n_5, n_6 >$ | Frequency (%) case (a) | case (b) |
|---|---|---|
| <0,2,8,4> | 5.86 | 2.55 |
| <0,3,6,4> | 4.64 | 2.87 |
| <0,1,10,2> | 4.34 | 3.29 |
| <0,3,6,5> | 3.44 | 1.07 |
| <0,2,8,5> | 2.78 | 0.72 |
| <0,1,10,3> | 2.52 | 1.39 |
| <0,3,6,6> | 2.39 | 0.40 |
| <0,4,4,6> | 2.35 | 0.74 |
| <0,1,8,4> | 0 | 3.03 |
| <0,2,6,4> | 0 | 2.55 |
| <0,2,6,5> | 0 | 2.27 |
| <0,1,8,3> | 0 | 2.05 |

To investigate the role of the small area faces in the distribution of the Voronoi-indexes a second calculation was made. In this investigation the faces with small area (smaller than the 'cutoff area-ratio') were identified, and the edge-numbers of these faces were simply excluded from the statistics (case (b) in the tables 4 and 5). The results are shown for 'cutoff area-ratio' = 1%. The originally most prominent Voronoi-polyhedra types were less frequent in the modified lists and new polyhedron-types appeared. This shows that several of these primarily prevalent polyhedra have at least one small-area face and the indexes of these cells are affected by distant 'nearest' neighbours.

*4.4. Investigation of common neighbours*

The packing structure can be further investigated by common neighbour analysis [20]. The first step in this analysis is the determination of the common neighbours of two neighbouring particles. The definition of neighbours can be different as was shown in the previous sections. The number of the common neighbours of two neighbouring particles depends on the definition of neighbours necessarily. Accordingly to the previous sections four neighbour definitions were used:

   Case 1: Voronoi nearest neighbours determined by the Voronoi tessellation method.

   Case 2: Voronoi neighbours with a common face's area higher than 1 % of the total surface area

('cutoff area-ratio' = 1%).

Case 3: Neighbours which are closer to each other than the cutoff-distances ($r_{max}$) determined from partial pair correlation functions: $r_{PdPd}$=3.5 Å, $r_{PdGe}$=3.6 Å, $r_{GeGe}$=3.6 Å.

Case 4: Similarly to Case 3 the pairs are determined by cutoff-distances, but in this case these distances are higher: $r_{PdPd}$=3.75 Å, $r_{PdGe}$=3.78 Å, $r_{GeGe}$=3.735 Å. These cutoff distances give the same coordination numbers as Case 2 (see section 4.2.).

The distributions of the most prominent common neighbour numbers (3, 4, 5 and 6) for the four neighbour definitions are collected in tables 6 and 7.

**Table 6.** Frequency distribution of the number of common neighbours around Ge atoms. Cases 1-4 denote different neighbour-definitions (see text).

| Number of common neighbours | Frequency (%) | | | |
|---|---|---|---|---|
| | Case 1 | Case 2 | Case 3 | Case 4 |
| 3 | 1.19 | 8.68 | 17.63 | 11.13 |
| 4 | 19.25 | 43.48 | 43.23 | 32.75 |
| 5 | 44.87 | 42.76 | 32.43 | 38.53 |
| 6 | 26.27 | 4.09 | 2.96 | 13.05 |

**Table 7.** Frequency distribution of the number of common neighbours around Pd atoms. Cases 1-4 denote different neighbour-definitions (see text).

| Number of common neighbours | Frequency (%) | | | |
|---|---|---|---|---|
| | Case 1 | Case 2 | Case 3 | Case 4 |
| 3 | 0.55 | 5.82 | 16.34 | 8.11 |
| 4 | 14.28 | 38.20 | 41.90 | 30.93 |
| 5 | 46.97 | 48.91 | 34.71 | 44.24 |
| 6 | 30.68 | 6.58 | 3.90 | 14.05 |

The number of common neighbours strongly depends on the neighbour-definition. In the case of Voronoi neighbours 5 common neighbours are the most prevalent both around Ge and Pd. In the case of the 'cutoff-distance method' the average number of common neighbours decreases, and the most frequent value is 4. The distributions of the common neighbour numbers in the Case 2 and Case 4 are more similar, because the average coordination numbers are the same (see section 4.2.), although there are still significant differences.

## 5. Conclusions

The structure of $Pd_{81}Ge_{19}$ metallic glass was investigated by neutron and X-ray diffraction and extended X-ray absorption fine structure spectroscopy. The experimental data sets were fitted simultaneously by the reverse Monte Carlo simulation technique. It was found that the average number of Ge atoms around Ge atoms is less than 0.3. Three-dimensional configurations compatible with the measurements were obtained without Ge-Ge nearest neighbours. The first neighbour shells of Ge and Pd atoms were determined by the cutoff distance method and the Voronoi tessellation method as well. On the average, 10.5-11 Pd atoms were found around Ge atoms. The average coordination number of Pd atoms was 12 by the cutoff distance method and 14 by the Voronoi tessellation method. The role of distant 'Voronoi nearest neighbours' in the average coordination numbers and in the numbers of common nearest neighbours was found to be significant.

**Acknowledgment**

I. P. and P. J. were supported by NKFIH (National Research, Development and Innovation Office) Grant No. SNN 116198. The neutron diffraction experiment was carried out at the ORPHÉE reactor, Laboratoire Léon Brillouin, CEA-Saclay, France.